\documentclass[aps,nofootinbib, prl, amsmath, floats, floatfix, twocolumn,superscriptaddress, showpacs,reprint]{revtex4}

\usepackage{graphicx}
\usepackage{amsmath,amssymb}
\usepackage{amsfonts}
\usepackage{xspace} 
\usepackage[usenames]{color}
\usepackage{dcolumn}
\usepackage{bm}
\usepackage{mathrsfs}
\usepackage[colorlinks=true]{hyperref}
\usepackage[all]{hypcap} 
 \usepackage[utf8]{inputenc}

\def\be{\begin{equation}}
\def\ee{\end{equation}}
\def\beq{\begin{eqnarray}}
\def\eeq{\end{eqnarray}}

\newcommand{\tn}{\textnormal}

\begin{document}

\title{Constraining Modified Theories of Gravity with Gravitational-Wave Stochastic Background}

\author{Andrea Maselli}\email{andrea.maselli@uni-tuebingen.de}
\affiliation{Theoretical Astrophysics, Eberhard Karls University of Tuebingen, Tuebingen 72076, Germany}
\author{Stefania Marassi}
\affiliation{INAF - Osservatorio Astronomico di Roma, Via di Frascati 33, I-00040 Monteporzio, Italy}
\author {Valeria Ferrari}
\affiliation{Dipartimento di Fisica, Sapienza Universita di Roma \& Sezione INFN Roma 1, P.A. Moro 5, 00185, Roma, Italy}
\author {Kostas Kokkotas}
\affiliation{Theoretical Astrophysics, Eberhard Karls University of Tuebingen, Tuebingen 72076, Germany} 
\affiliation{Center for Relativistic Astrophysics and School of Physics, Georgia Institute of Technology, Atlanta, Georgia 30332, USA} 
\author{Raffaella Schneider}
\affiliation{INAF - Osservatorio Astronomico di Roma, Via di Frascati 33, I-00040 Monteporzio, Italy}

\pacs{04.30.-w, 04.80.Cc, 04.50.Kd}

\date{\today}
\begin{abstract}

The direct discovery of gravitational waves has finally opened a new
observational window on our Universe, suggesting that the population of
coalescing binary black holes is larger than previously expected. These
sources produce an unresolved background of gravitational waves,
potentially observable by ground-based interferometers. In this Letter we investigate how
modified theories of gravity, modeled using the parametrized post-Einsteinian formalism, 
affect the expected signal, and analyze the detectability of the resulting stochastic
background by current and future ground-based interferometers.  We
find the constraints that Advanced LIGO would be able to set on
modified theories, showing that they may significantly
improve the current bounds obtained from
astrophysical observations of binary pulsars. 
\end{abstract}

\maketitle 

\noindent{\bf{\em I. Introduction.---}}
The LIGO and Virgo Collaborations has recently announced the first direct detection of gravitational 
waves (GW) \cite{Abbott:2016blz}. The loudness of the GW 150914 event, with an unexpected 
high signal-to-noise ratio (SNR), has allowed us to associate GW 150914 to the coalescence of a binary 
black hole (BBH) system with (source-frame) masses $\rm 36_{-4}^{+5}M_\odot$ and 
$\rm 29_{-4}^{+4}M_{\odot}$ at a luminosity distance of $\rm \sim 400 Mpc$. This binary 
detection implies BBH masses and coalescence rates higher than previous theoretical 
predictions \cite{Abadie:2010cf,2015MNRAS.451.4086S}, and in agreement with recent 
estimates, obtained with a population synthesis approach, which predicts the formation of 
a detectable BBH in the early Universe, and in low metallicity environments \cite{Dominik:2012kk,Dominik:2014yma}.

As a consequence, also the stochastic gravitational wave background (GWB) produced by 
these coalescing cosmological BBH sources should be at the higher end of previous estimates 
\cite{Allen:1997ad,Zhu:2011bd,PhysRevD.84.084004,1674-4527-11-4-001,PhysRevD.84.124037,PhysRevD.85.104024}, 
and could be potentially detectable by advanced detectors 
\cite{TheLIGOScientific:2016wyq}.  

In this Letter we explore for the first time the ability of terrestrial interferometers to 
constrain the fundamental parameters  of modified theories of gravity, through
the detection of the GWB generated by the coalescing BBH.
To this aim, we compare the {\itshape fiducial} GWB computed assuming general relativity (GR) 
\cite{TheLIGOScientific:2016wyq} with the signal produced by modified theories. We model 
deviations from GR using the parametrized post-Einsteinian (PPE) formalism 
\cite{PhysRevD.80.122003}, which has been developed to capture GR modifications in the 
GW data. More specifically, similar to the post-Newtonian (PN) formalism, which is a 
low-velocity and weak-field expansion of the metric and matter variables \cite{lrr-2014-2},
the PPE approach traces back model independent deviations from GR directly into GW 
templates. The relevance of such corrections in the gravitational emission of compact binaries 
has been deeply investigated in the literature  \cite{Yunes:2016jcc,Narikawa:2016uwr,PhysRevD.87.102002,PhysRevD.86.022004,PhysRevD.87.102001,PhysRevD.88.064056,PhysRevD.88.089902}, 
showing that GW parameter estimation can be strongly affected by GR deviations, 
if they are not properly taken into account \cite{Vitale:2013bma}. Astrophysical
constraints on the lowest order PPE coefficients have been set using
observations of relativistic binary pulsars \cite{PhysRevD.82.082002}. We refer
the reader to the seminal manuscript \cite{PhysRevD.80.122003}, and to the
review papers \cite{Yunes:2013dva,Berti:2015itd} (and references therein), for
an exhaustive description of this topic.
 
Following the PPE approach, in this work we do not focus on any specific theory
of gravity.  Conversely, we carry out a completely {\it agnostic} analysis. We
investigate the regions of the PPE parameter space that are  more likely to
contribute to the GWB, producing significant deviations from GR. We consider
both second and third generation detectors,
analyzing their ability to detect the modified signal and to extract its
physical properties. Finally, we show how terrestrial GW interferometers can
improve the bounds on the  PPE coefficients  set by binary pulsar observations
\cite{PhysRevD.82.082002}.

\noindent{\bf{\em II. The stochastic background.---}}
The normalized GWB spectral energy density is defined as
\begin{equation}
\Omega_\tn{GW}(f)= \frac{1}{\rho_c}\frac{\rm d\rho_\tn{GW}}{d\ln f}, 
\end{equation}
where $\rho_\tn{GW}$ is the GW energy density and $\rm \rho_{c}=\frac{\rm 3H_{0}^{2}}{\rm 8\pi G}$ 
is the critical energy density required to close the Universe. We assume 
$\rm H_{0} = 70$km/s/Mpc, $\rm \Omega_{m}=0.3$, $\rm \Omega_{\Lambda}=0.7$. 
The function $\Omega_\tn{GW}$  can also be written as
\begin{equation}
\Omega_\tn{GW}(f)=\frac{f}{\rm H_0\rho_c}\int \frac{ \frac{dE_\tn{GW}}{df}[f(1+z)]{\cal R}_{coal}(z)}{(1+z)E(z)}dz,
\end{equation}
where $\frac{\rm dE_{GW}}{df}$ is the rest-frame GW spectrum emitted by a single 
source, $E(z)=\sqrt{\rm \Omega_{m}(1+z)^3+\Omega_{\Lambda}}$ for 
a flat Universe, and ${\cal R}_{coal}(z)$ is the BBH observed event rate per comoving volume. 
Following \cite{TheLIGOScientific:2016wyq}, we assume as the {\itshape fiducial} model ${\cal
R}_{coal}(z)$ proportional to the cosmic star formation rate
\cite{Vangioni:2014axa,Kistler:2013jza}, weighted by the fraction of stars with
metallicity $Z<0.5 Z_{\odot}$ (see APPEndix B in \cite{Callister:2016ewt}).

\noindent{\bf {\em III. The waveform model.---}}
In the PPE approach the gravitational waveform in the frequency domain is modified both in 
amplitude and phase with respect to the PN waveforms,
\begin{equation}
h(f)=h_\tn{GR}(f)(1+\alpha u^\beta)e^{i\delta u^\zeta}\ ,\label{PPEwave}
\end{equation}
where $u=(\pi{\cal M} f)^{1/3}$, ${\cal M}$ is the chirp mass of the system, and 
$(\alpha,\beta,\delta,\zeta)$ are PPE parameters. $\beta$ and $\zeta$ define the type 
of modification introduced in the theory, while $\alpha$ and $\delta$ control the 
magnitude of the deviation, and have to be constrained by data. 
(In \cite{Yunes:2016jcc} Yunes and collaborators have shown the
equivalence between the framework used by the LIGO Collaboration to test GR 
in \cite{PhysRevLett.116.221101}, and the PPE formalism adopted in our paper.
As pointed out in \cite{Yunes:2016jcc}, the waveforms employed by the LIGO
Collaboration represent a subset of the PPE models, since they allow
only positive GR modifications in the phase.) 
In this Letter, we focus on the information which can be extracted from the wave
amplitude, therefore on $\alpha$ and $\beta$. $h_\tn{GR}(f)$ is the
phenomenological waveform described in \cite{Ajith:2007kx}, which combines the
PN approximation with numerical relativity results to describe the whole binary
coalescence.  Moreover, we only consider the inspiral part of the signal,
deferring to a forthcoming paper a detailed analysis of the impact of GR
modifications on the merger and ringdown phases \cite{paper2}. For
$\alpha=\delta=0$ we recover the standard PN waveform which, at lowest order,
has amplitude \begin{equation} {\cal A}(f)=\sqrt{\frac{5}{24}}\frac{{\cal
M}^{5/6}}{\pi^{2/3}d}f^{-7/6}\ , \end{equation} where $d$ is the source
distance. In our analysis we truncate the template at the merger frequency in
GR, which can be parametrized in terms of the mass components of the binary
\cite{Ajith:2007kx}. The waveform \eqref{PPEwave} enters quadratically into the
GWB through the GW single source spectrum $dE_\tn{GW}/df$, i.e.,
$\Omega_\tn{GW}(f)\propto h_\tn{GR}(f)^2(1+\alpha u^\beta)^2$.

As noted in \cite{PhysRevD.80.122003}, Eq.~\eqref{PPEwave} does not describe the 
most general modified waveform, and can be thought of as a single-parameter deformation 
of GR. Although multiple PPE coefficients may enter into the gravitational signals of a 
certain theory of gravity, the templates we consider parametrize the effects that are 
more relevant in the interferometer's bandwidth. It should be mentioned that a feature of 
this approach is that the map between the PPE parameters and a specific theory is not 
unique; thus, there could be more than one model yielding the same result 
\cite{Yunes:2016jcc}. However, the detection of these coefficients would provide 
precious information on the theory of gravity. For example, a measurement of $\alpha\neq0$ 
for $\beta=1$ would identify a parity violation, while for $\beta=-8$ it would be a hint 
of anomalous acceleration, or violation of position invariance \cite{Yunes:2013dva,Yunes:2016jcc}.

In this Letter we do not choose any particular modified theory  of gravity. Rather, 
being completely agnostic on the real nature of gravity, we  explore the PPE 
parameter space to study how the modified waveform affects the GWB produced by the BBH. 
However, we  assume the GR corrections in Eq.~\ref{PPEwave} as 
perturbative terms, and accordingly, in our analysis we consistently consider 
values of $\alpha$ and $\beta$ that satisfy the bound 
$\vert\alpha u^{\beta}\vert<1$.

The PPE parameters have already been constrained by astrophysical observations.
Using the data of double binary pulsars, strong bounds have been set 
on the amplitude $\alpha$ as a function of the specific considered theory,
identified by the exponent $\beta$  \cite{PhysRevD.82.082002}. 
It is shown  that negative values of $\beta$ yield very tight constraints on 
$\alpha$. For gravity theories with $\beta=-2$, which gives 
$-1$ PN corrections in the amplitude, these observations imply 
$\vert \alpha\vert\lesssim 10^{-9}$. In general, theories with $\beta<0$ predict 
corrections to GR that affect the low frequency regime and therefore, 
are well constrained by electromagnetic observations of binary systems far from 
coalescence, or by future GW space interferometers \cite{Barausse:2016eii}. 
In this Letter we will focus our analysis on modified waveforms with $\beta>0$.  
In these models the gravitational waveforms exhibit corrections at higher frequencies,
and therefore are ideal candidates to be tested in the near future by ground-based 
GW interferometers. 

To quantify the differences between the GR and the modified background, we 
introduce the optimized SNR for a given integration time 
$\rm T$ \cite{Allen:1997ad}:
\begin{equation}\label{SNR}
\tn{SNR}=\frac{3H^2_0}{\sqrt{50}\pi^2}\sqrt{\rm T}\int_{0}^{\infty}df 
\left[\frac{\gamma^2(f)\Omega_\tn{GW}^2(f)}{f^6S_1(f) S_2(f)}\right]^{1/2}\ ,
\end{equation}
where $S_{1}(f),S_2(f)$ are the power spectral noise densities of two detectors, and 
$\gamma(f)$ is the normalized overlap reduction function. 
We have computed the SNR for Advanced LIGO (AdLIGO) and the Einstein
Telescope (ET), assuming the \texttt{ZERO\_DET\_high\_P} anticipated sensitivity for both Livingston and
Hanford sites \cite{LIGOnoise}, and the ET-B configuration for the 
\cite{Howell:2010hm}.  For the latter, $\gamma(f)$ is assumed to be
constant, i.e., $\gamma=-3/8$, while for AdLIGO the overlap function is given by
the numerical results described in \cite{PhysRevD.88.124032}. 

\begin{figure}[ht]
\includegraphics[width=8cm]{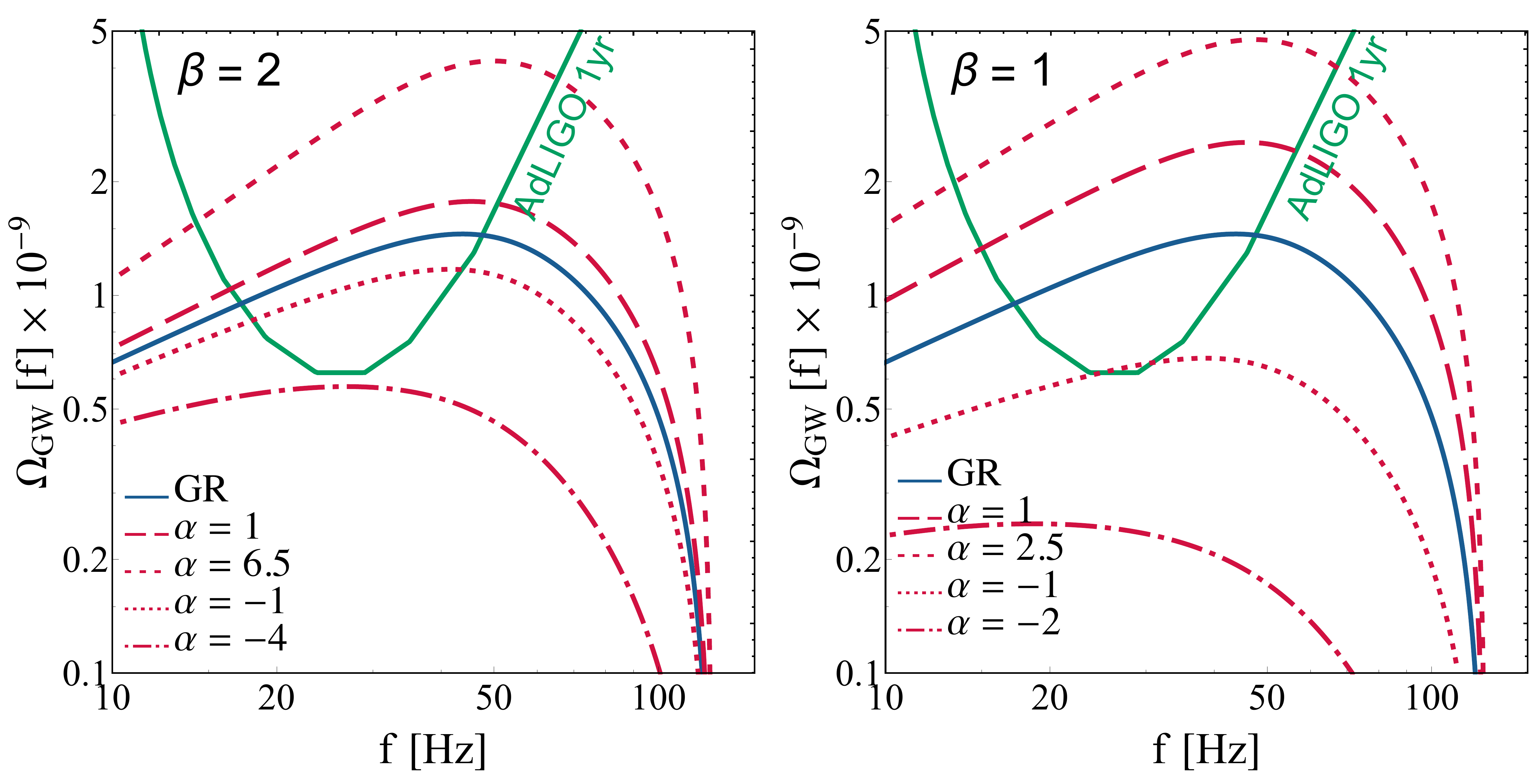}
\caption{The spectral energy density $\Omega_\tn{GW}(f)$ is plotted as a
function of frequency. $\Omega_\tn{GW}(f)$ is computed in GR and in modified
theories with PPE parameter $\beta=2$, and $\beta=1$, and different values of 
$\alpha$. The power-law integrated sensitivity curve for 1 year of integration 
with AdLIGO is also shown.}
\label{fig:omegaf1} \end{figure}

\noindent{\bf{\em IV. Results.---}}
In Fig.~\ref{fig:omegaf1} we show the GWB spectra for modified theories with exponent 
$\beta=(2,1)$ and different values of the parameter $\alpha$, compared to the {\it
fiducial} GR case, (We note that the {\it fiducial} model considered in this work 
differs from the GWB of \cite{TheLIGOScientific:2016wyq}, which is computed also taking 
into account the merger and the ringdown phases.) and assuming a mean chirp mass of 
${\cal M}=28M_\odot$. The power-law integrated sensitivity curve for AdLIGO with an integration 
time of 1 year is also shown.

The net effect of positive (negative) values of $\alpha$ is to increase (decrease) the 
spectral energy density of the background, which, for certain values, is significantly 
different from that predicted by GR. As an example, a gravity theory with $\beta=2$, 
which yields a 1 PN correction to the amplitude of the waveform, and  $\alpha \sim 6$, 
would produce a background three times larger than the {\it fiducial}.
A similar behaviour is shown for GWBs with $\beta=1$. As expected by 
the PN character of the PPE approach, for a fixed $\alpha$, smaller values 
of the exponent $\beta$ yields larger deviations.

\begin{figure}[th] 
\includegraphics[scale=2,width=8cm]{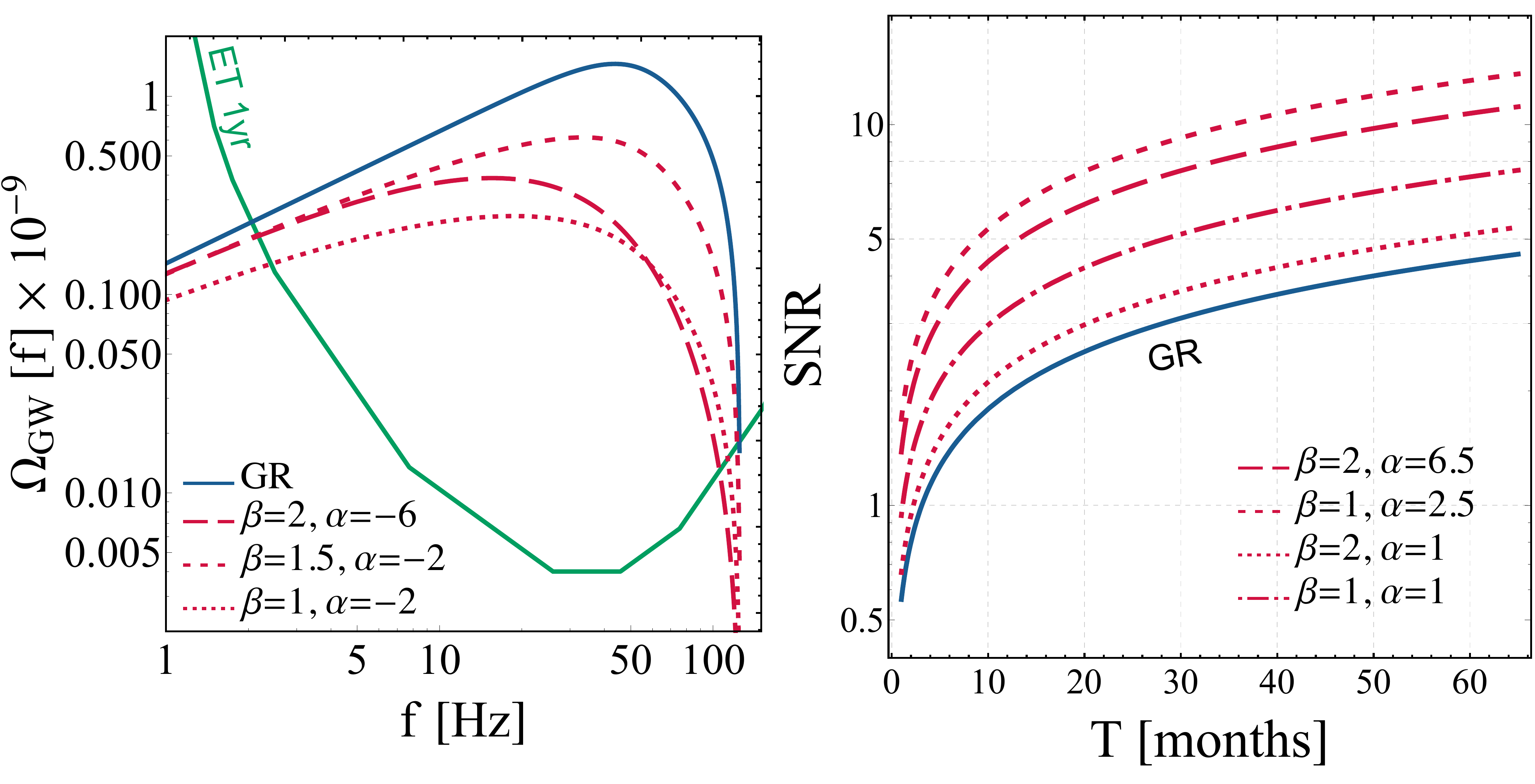}
\caption{(Left) Same as Fig.~\eqref{fig:omegaf1}, for values of negative
$\alpha$ and different values of $\beta$, compared to the ET
sensitivity curve for one year of observation. (Right) 
The SNR with which AdLIGO would detect a PPE background is plotted
for some of the gravity theories considered in Sec.~IV versus the integration
time.}
\label{fig:omegaf2} \end{figure}

When $\alpha<0$ the amplitude of the GWB decreases, limiting the 
possibility to detect these backgrounds with advanced detectors.
However, they are potentially observable by a third 
generation of ground based interferometers. The left panel of 
 Fig.~\ref{fig:omegaf2} shows the GWB for some values of
$\alpha<0$ and $\beta=(2,1.5,1)$, compared to the power-law integrated
sensitivity curve of ET, assuming 1 year of observation. 

In the right panel of Fig.~\ref{fig:omegaf2} we show how the SNR changes
as a function of the integration time, for AdLIGO and different PPE models.
For some of the considered configurations the SNR increases to a factor
$\gtrsim 10$ after 24 months. The fiducial GR background would require $\sim$30
years to reach the same value.
\begin{figure}[th] \includegraphics[width=8cm]{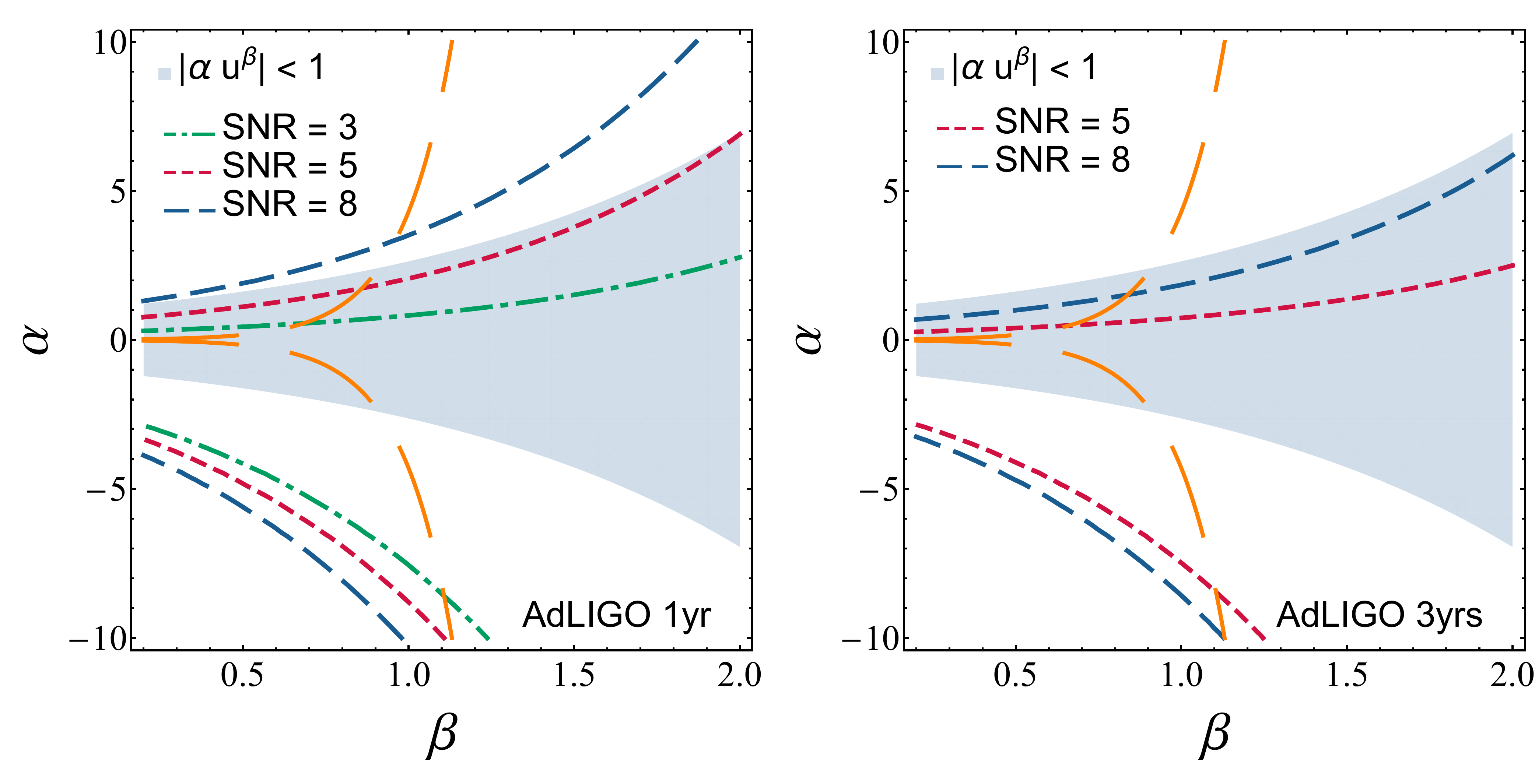} \caption{Contour lines
corresponding to different SNR, for gravity theories with PPE parameters
$\alpha>0$ and $\beta\in[0.2,2]$ assuming 1 and 3 years of integration with
AdLIGO. The shaded region identifies the parameter space where the PPE
parameters satisfy the bound $\vert\alpha u^{\beta}\vert<1$, 
while the long-dashed curves correspond to the
pulsar constraint \cite{PhysRevD.82.082002}. The allowed parameter
space is the colored region on the right side of the pulsar
constraint.} \label{fig:SNR} \end{figure}

In Fig.~\ref{fig:SNR} we extend our analysis, showing the contour lines for
detection thresholds SNR$=(3,5,8)$, computed for
AdLIGO with 1 and 3 years of integration, for theories with $\alpha>0$
and $\beta\in[0.2,2]$. The long-dashed curve identifies the region where the
parameters $\alpha$ and $\beta$ are constrained by binary pulsar observations,
as computed in \cite{PhysRevD.82.082002}: the allowed region is on the right of
this curve.
The shaded region defines the range where $\alpha$ and $\beta$ satisfy the
condition $\vert \alpha u^{\beta}\vert<1$.
After 1 year of integration, AdLIGO would be able
to identify GWBs with SNR = 5, produced by modified theories with
$\beta\gtrsim0.9$ and values of $\alpha$ lying on the red dashed curve.
Three years of observation would be needed to detect the same signals with 
$8\lesssim \tn{SNR}\lesssim10$.
 
In Table~\ref{tableSNR} we show the SNR computed for the advanced and third 
generation interferometers, for different integration times, for the GWB computed 
using the PPE waveforms  with $\beta=2$. Large SNRs are expected for 
ET, (We note that such SNRs may be biased since Eq.~\eqref{SNR} is defined 
in the small signal approximation, whereas ET should be able to detect these 
backgrounds directly.) but for some values of $\alpha$ and $\beta$ the GWB 
could be potentially detectable also by AdLIGO. 
\begin{table}[ht] \centering \begin{tabular}{ccccc} \hline \hline $\beta=2$ &
AdLIGO  & AdLIGO & ET\\ $\alpha$& (1 yr) & (3 yrs)  & (1 yr) \\ \hline
\textnormal{GR} &  1.39 &  1.96  &  223 \\ 1 & 2.31 & 4.00 & 656 \\ 4 & 3.54 &
6.13 & 1060 \\ 6.5 & 4.78 & 8.28 & 1470 \\ -1 & 1.64 & 2.85 & 446 \\ -4 & 0.887
& 1.54 & 216 \\ -6.5 & 0.47 & 0.814 & 102 \\ \hline \hline \end{tabular}
\caption{SNR of AdLIGO and ET computed for different integration times,
$\beta=2$ and different values of $\alpha$.} \label{tableSNR} \end{table}
The analysis presented above shows that a region of the PPE parameter space does
exist, where the spectral energy density $\Omega_\tn{GW}(f)$ of the GWB produced
by binary black hole coalescence could be detected by AdLIGO. To further clarify
this point, we assess the ability of current interferometers to distinguish
these GWB from the GR counterpart, and to extract physical information. We 
follow the strategy adopted in \cite{Callister:2016ewt}, where it has recently
been pointed out that second generation detectors may not be able to distinguish
between a BBH GWB and a generic power-law background. This would strongly affect
our ability to extract information on the background shape. Here, we apply a
model selection procedure to determine whether the modified GWB can be discerned
by  one computed in GR, or assuming a power-law behavior.  To this aim we
compare the likelihood functions between two models $\Omega_{1,2}(f)$: $ {\cal
L}(\Omega_1\vert\Omega_2)\propto\tn{Exp}\left[-\frac{1}{4}(\Omega_1-\Omega_2\vert\Omega_1-\Omega_2)\right]\
, $ where $ (A\vert
B)=2T\left(\frac{3H_0^2}{10\pi^2}\right)^2\int_{f_\tn{min}}^\infty
df\gamma^2(f)\frac{A(f)B(f)}{f^6 S_1(f)S_2(f)}$. 

\begin{figure}[th]
\includegraphics[width=8cm]{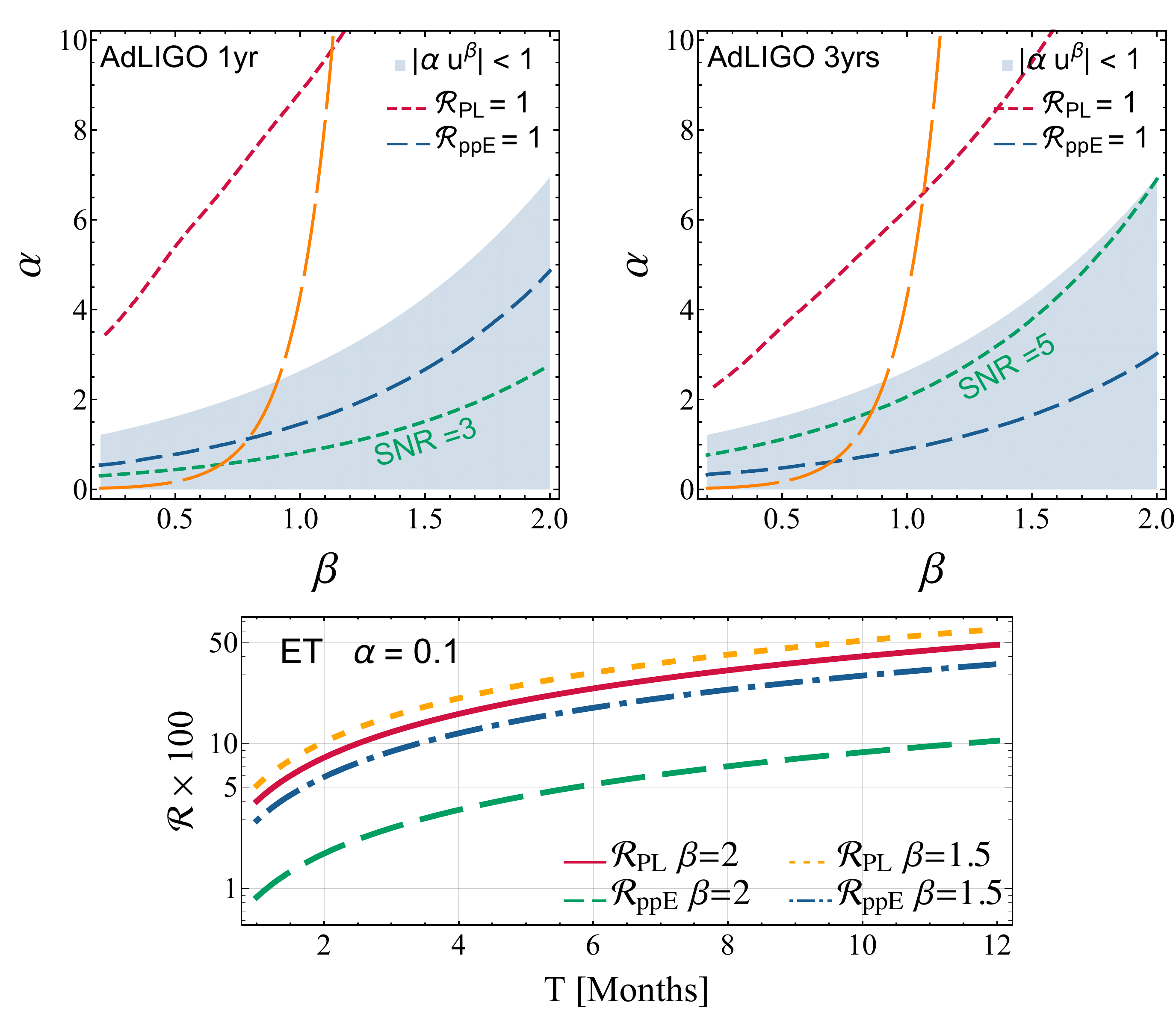}
\caption{(Top) Values of $\alpha$ and $\beta$ yielding ${\cal R}_\tn{PL}=1$
and ${\cal R}_\tn{PPE}=1$, computed for AdLIGO with 1 and 3 years of
integration. The long-dashed line and the shaded region correspond to the pulsar
constraints and the parameter space where $\vert\alpha u^{\beta}\vert<1$,
respectively. The green curves represent the contour line for the GWB detectable
with SNR =3 and 5.  (Bottom) Likelihood ratios ${\cal R}_\tn{PPE}$ and
${\cal R}_\tn{PL}$, computed for ET as a function of
the integration time, for PPE models with $\alpha=0.1$ and $\beta=(2,1.5)$.}
\label{fig:like} \end{figure}

Then, we compute their likelihood ratios,
\begin{equation}\label{likeratio} {\cal R}_\tn{PPE}=\ln\frac{{\cal
L}(\Omega_\tn{GR}\vert\Omega_\tn{GR})}{{\cal
L}(\Omega_\tn{GR}\vert\Omega_\tn{PPE})}\ \ ,\ {\cal R}_\tn{PL}=\ln\frac{{\cal
L}(\Omega_\tn{PPE}\vert\Omega_\tn{PPE})}{{\cal
L}(\Omega_\tn{PPE}\vert\Omega_\tn{PL})}\ , \end{equation} where
$\Omega_\tn{PPE}=\Omega_\tn{PPE}(\alpha,\beta)$ and as usual
$\Omega_\tn{GR}=\Omega_\tn{PPE}(\alpha=0)$, while the power-law density is given
by $\Omega_\tn{PL}=\Omega_0(f/f_0)^{2/3}$, with $f_0$ being the arbitrary reference
frequency and $\Omega_0$ the amplitude that can be computed analytically
\cite{Callister:2016ewt}. If the likelihood ratio approaches 1, the two GWBs
cannot be distinguished, while large values of ${\cal R}$ identify a preferred
model. In particular ${\cal R}_\tn{PPE}\gg1$ suggests that
$\Omega_\tn{PPE}\neq\Omega_\tn{GR}$, and ${\cal R}_\tn{PL}\gg1$ reveals 
that the detected GWB differs significantly from a power-law energy spectrum. To
assess the full detectability of the features of the GW signal, both ratios in
Eq.~\eqref{likeratio} must be greater than 1. The top panels of
Fig.~\ref{fig:like} show the values of $\alpha$ and $\beta$ for which ${\cal
R}_\tn{PPE}=1$ and ${\cal R}_\tn{PL}=1$, with 1 and 3 years of integration
with AdLIGO, compared against the binary pulsar constraints (long-dashed curve).

We note that for $\beta\gtrsim1$ all configurations in the allowed region
(shaded zone on the right of the long-dashed curve) lead to ${\cal
R}_\tn{PPE}>1$, and then can potentially be distinguished from the {\it
fiducial} model, with a cumulative SNR $\lesssim4$. Three years of observation
would improve this picture, allowing us to discern among gravity theories
with ${\cal R}_\tn{PPE}>>1$ and SNR $\gtrsim5$. However, larger values of the
PPE amplitude, outside the permitted parameter space, are needed to satisfy the
condition ${\cal R}_\tn{PL}\gg1$.  For example, a gravity theory with
$\beta=1.5$ requires $\alpha \gtrsim 12$.

Third generation detectors, would be able to fully extract the 
physical information of the GWB in the allowed parameter space, and constrain
the change in slope in Figs.~\ref{fig:omegaf1}-\ref{fig:omegaf2} due to the PPE
correction, which is frequency dependent. 
In the bottom panel of Fig.~\ref{fig:like} we show
${\cal R}_\tn{PPE}$ and ${\cal R}_\tn{PL}$ for two PPE theories with
$\alpha=0.1$ and $\beta=(1.5,2)$, as a function of the integration time, for
ET. For both theories we find
high values of both likelihood ratios, even after 6 months of
observation.

It is interesting to note that even a {\it null detection} of the GW signal
would provide  information on the allowed space for the PPE parameters. We
propose here a simple strategy to exploit this feature. As a rule of thumb
we can assume that the GWB is potentially observable if the SNR is greater than
a defined threshold SNR$_\tn{T}$. Looking at 
Fig.~\ref{fig:SNR}, after 1 year of integration and assuming SNR$_\tn{T}=3$,
if no GWB is detected, we could exclude the parameter space outside the two
green dot-dashed curves. A SNR threshold of 5 rules out
the values of $\alpha$ and $\beta$ outside the red
short-dashed curves. Two or more years of observation would provide additional
restrictions on the PPE coefficients. This simple strategy would constrain the
PPE amplitude $\alpha$ to values $\mathcal{O}(10)$, with a large impact on
models with $\beta \gtrsim 1 $, where current bounds are quite loose.  In fact,
for a gravity theory with $\beta =2$, binary pulsar observations can only
constrain $\alpha$ to be $\vert\alpha\vert\lesssim2000$. Our approach would
improve this bound by 2 orders of magnitude.

\noindent{\bf {\em IV. Conclusions.---}}
In this paper we have analyzed how GR modifications affect the GWB produced by the 
coalescence of BBH systems, showing that the parameter space available for modified 
theories may yield an enhancement of the background energy density. 

As pointed out in \cite{TheLIGOScientific:2016wyq} the fiducial GWB  has an
uncertainty band which depends on different assumptions on the formation and
evolution of the binary progenitors.  Alternative theories introduce another
source of degeneracy.  However, for a given astrophysical scenario, every GWB
computed in GR has a modified  PPE counterpart that, for $\alpha>0$, is larger
in amplitude and different in slope.  These features imply that a GWB detection
will still be able to constrain the PPE parameters in large regions of the
parameter space. 
In addition, it should be mentioned that in the future more detectors, 
Virgo \cite{TheVirgo:2014hva}, KAGRA \cite{Aso:2013eba} and LIGO-India \cite{LIGOIndia}, 
will become operational; the sensitivity of the  network of these detectors 
will significantly increase with respect to that of AdLIGO alone, making accessible 
further regions of the PPE parameter space. 

A detailed investigation of the regions of $\alpha$ and $\beta$ where the GWB is
distinguishable from the GR background requires a more sophisticated
statistical analysis, like the one presented in \cite{PhysRevLett.109.171102}. We
plan to include this analysis in a forthcoming extended publication \cite{paper2},
in which we will also consider the effect of a network of detectors. 
Finally, we remark that a  comprehensive
study of how the merger and the ringdown may affect the GWB in GR was carried
out in \cite{PhysRevD.84.124037,Zhu:2011bd}, pointing out that only ET would be able
to identify the contribution of these two phases.  It would be interesting to
reexamine these results for AdLIGO, as far as modified theories of gravity are
considered.

\noindent{\em Acknowledgements.---}
The research leading to these results has received funding from the European Research 
Council under the European Union's Seventh Framework Program (FP/2007-2013)/ERC 
Grant No. 306476. It is a pleasure to thank Nico Yunes, Emanuele Berti and Paolo 
Pani for having carefully read the manuscript and for their useful comments.



\bibliography{bibnote}

\end{document}